\documentclass[preprint,12pt]{imsart}
\usepackage{mathtools}
\RequirePackage[OT1]{fontenc}
\usepackage{natbib}
 \usepackage{graphicx}
 \usepackage{mhchem}
 \usepackage{multirow}
 \usepackage{amssymb}

    \usepackage{amsmath,amssymb,euscript,amsthm,amsfonts,mathrsfs,amscd}
\usepackage{color}
    \usepackage{amssymb,euscript,amsfonts,mathrsfs,amscd}
\usepackage{amsmath}

\usepackage{algorithmic,algorithm2e,float}

\usepackage{enumitem}

\startlocaldefs
\endlocaldefs

\begin{document}

\begin{frontmatter}

\title{Improving Safety of the Continual Reassessment Method via a Modified Allocation Rule}
\runtitle{A new allocation rule for Phase I clinical trials}

\author{\fnms{Pavel} \snm{Mozgunov}\ead[label=e1]{p.mozgunov@lancaster.ac.uk}}
\and
\author{\fnms{Thomas} \snm{Jaki}\ead[label=e2]{t.jaki@lancaster.ac.uk}}

\runauthor{P. Mozgunov and T. Jaki}

\affiliation{Lancaster University}

\address{Department of Mathematics and Statistics, \\ Lancaster University, Lancaster, LA1 4YF, UK  \\ \printead{e1}}

\begin{abstract}
This paper proposes a novel criterion for the allocation of patients in Phase~I dose-escalation clinical trials aiming to find the maximum tolerated dose (MTD). Conventionally, using a model-based approach the next patient is allocated to the dose with the toxicity estimate closest (in terms of the absolute or squared distance) to the maximum acceptable toxicity. This approach, however, ignores the uncertainty in point estimates and ethical concerns of assigning a lot of patients to overly toxic doses. Motivated by recent discussions in the theory of estimation in restricted parameter spaces, we propose a criterion which accounts for both of these issues. The criterion requires a specification of one additional parameter only which has a simple and intuitive interpretation. We incorporate the proposed criterion into the one-parameter Bayesian continual reassessment method (CRM) and show, using simulations, that it results in the same proportion of correct selections on average as the original design, but in fewer mean number of toxic responses. A comparison to other model-based dose-escalation designs demonstrates that the proposed design can result in either the same mean accuracy as alternatives but fewer number of toxic responses, or in a higher mean accuracy but the same number of toxic responses. We conclude that the new criterion makes the existing model-based designs more ethical without losing efficiency in the context of Phase I clinical trials.
\end{abstract}


\end{frontmatter}

\section{Introduction}
Consider a Phase I clinical trial with two doses ($d_1$,  $d_2$) and a binary endpoint, dose-limiting toxicity (DLT) or no DLT, to stress the importance of an allocation criterion. The goal of the trial is to find the maximum tolerated dose (MTD) which has a probability of a DLT closest to the pre-specified target, say $\gamma=0.30$. Assume that $10$ patients were assigned to each dose and $2$ and $4$ toxicities are observed, respectively. Then, a typical question in a sequential trial is: ``which dose should be administered to the next patient''. A conventional criterion for many dose-escalation model-based designs \cite[see e.g.][]{QPF90,wages2011} is to assign the next patient to dose $d_i$ corresponding to the point estimate  $\hat{p}_i$ closest to $\gamma$ in terms of the absolute or, equivalently, the squared distance
\begin{equation}
\left(\hat{p}_i-\gamma \right)^2.
\label{sq_distance}
\end{equation}
\noindent Assume that in the example above, the probabilities $p_1$ and $p_2$ are considered as random variables with Beta distributions $\mathcal{B}(2,8) $ and $\mathcal{B} (4,6)$ and one uses the mean as the point estimate: $\hat{p}_1=0.2$ and $\hat{p}_2=0.4$. Following criterion~(\ref{sq_distance}), the next patient can be allocated to either of doses as both estimates are equally close to the target. At the same time, one can argue that these doses are not ``equal'' for at least two reasons. On the one hand, the criterion~(\ref{sq_distance}) ignores the randomness of the estimates. Indeed, the probability of being within 5\% of $\gamma$ is larger for $p_2$
 \begin{equation}
\mathbb{P}\left(p_2 \in (0.25,0.35) \right) > \mathbb{P}\left(p_1 \in (0.25,0.35) \right).
\label{probabilities}
\end{equation}
The larger variance of $p_2$ favours the decision to allocate the next patient to $d_2$. On the other hand, the allocation of a patient to the dose with estimated toxicity probability of $0.4$ can be considered to be unethical as it exposes a patient to an unacceptably high toxicity. Moreover, it is usually of interest to balance these two aims in a Phase I clinical trial - an investigator would like to study more doses but would like to avoid exposing patients to doses far from the MTD. In the illustration above, the criterion~(\ref{sq_distance}) fails to address both of these concerns. 

The question of safety was firstly addressed by~\cite{ewoc} using the Escalation with Overdose Control (EWOC) design. The EWOC uses the criterion 
\begin{equation}
\mathbb{E}\left(\alpha (\gamma-p_i)^{+} + (1-\alpha) (p_i-\gamma)^{+} \right)
\label{ewoc_distance}
\end{equation}
for patients allocations \citep{cheung2011}, where $(x)^{+}=\max(0,x)$ and $\alpha$ is a parameter of asymmetry. The criterion~(\ref{ewoc_distance}) imposes that the allocation to a more toxic dose should have a more severe penalty than to a less toxic one. The EWOC design has been shown to result in a low average number of DLTs. However, it also leads to underestimation of the MTD in many realistic scenarios and some modifications were recently proposed by \cite{mourad2010,wheeler2017}. Another design aimed to resolve both the uncertainty and safety concerns is the Bayesian Logistic Regression Model (BLRM) by~\cite{neuenschwander2008}. To address the first concern, it is proposed to use the whole distribution of the DLT probability, while the ethical one is addressed via a penalty to overly toxic intervals. The allocation is determined by a loss function computed for each dose. While this approach has been proven to be useful in practice, it requires specifying several parameters (values of the loss function and corresponding intervals) on which the operating characteristics of the design can depend on significantly.

In this work, we propose a new criterion for the allocation of patients in dose-escalation trials. A point estimate of the criterion takes both the variance of the distribution of probability toxicity and the ethical concerns of an investigator into account. The novel criterion requires only one additional parameter, which has a simple and intuitive interpretation, to be specified. This parameter controls the trade-off between the uncertainty in estimates and the conservative of an investigator (in terms of the mean number of toxic responses). The novel criterion is defined as the squared distance generalised to a parameter defined on the restricted parameter space (the unit interval) as proposed by~\cite{mozgunov2018,mozgunov2017loss}. As it is generally agreed that model-based Phase I designs lead to better operating characteristics than rule-based alternatives~\citep{reiner1999}, we incorporate the proposed criterion into the Bayesian continual reassessment method \cite[CRM]{QPF90} and compare its operating characteristics to the traditional one-parameter power model CRM design, the EWOC design and its recent modifications, and the BLRM design.

Importantly, we will focus on the application of the novel criterion to the one-parameter power model as it has been shown~\citep{o1996likelihood,zohar2001,cheung2011,iasonos2016} to be able to identify the MTD with a high probability. Note, however, that the proposed criterion is generic and can be applied to any parametric model (for instance, the two-parameter logistic model) if it is preferred by an investigator.


The rest of the paper proceeds as follows. The new criterion and its properties are studied in Section 2. The application of the novel criterion in the context of an actual clinical trial is considered in Section 3. A simulation comparison to the traditional continual reassessment method is given in Section 4. A comparison to the EWOC-type and BLRM designs are given in Section 5. Section 6 concludes with a discussion.

\section{Methods}
\subsection{Criterion}
Consider a Phase I clinical trial with binary DLT outcomes and $m$ doses $d_1,\ldots,d_m$. The main estimation objective in a such Phase I trial is the probability of DLT $p_i \in (0,1)$ if dose $d_i$ was given to a patient. Once estimates of $p_i$ are obtained, an investigator selects the MTD as the dose associated with the toxicity probability closest to $\gamma \in (0,1)$. Let us consider the criterion~(\ref{sq_distance}) for some fixed dose $d$ and the associated probability $p$.  It is argued by~\cite{ait}, that the criterion~(\ref{sq_distance}) might not be a reliable measure of distance between objects defined on restricted parameter spaces. This argument is valid in the considered setting as both $p$ and $\gamma$ are defined on the restricted space - the unit interval. As an alternative, \cite{ait} proposed a new distance between $p$ and $\gamma$
\begin{equation}
AD(p,\gamma)= \sqrt{\left(\log \frac{p}{1-p} - \log \frac{\gamma}{1-\gamma} \right)^2}, \ p,\gamma \in (0,1)
\label{ait}
\end{equation}
known as the Aitchison distance. However, the Aitchison distance lacks some important properties such as convexity and a closed form solution for the corresponding minimizer~\citep{mozgunov2017loss}. Instead, \textit{the convex unit interval symmetric distance} 
\begin{equation}
\delta(p,\gamma) = \frac{(p-\gamma)^2}{p(1-p)}.
\label{criterion}
\end{equation}
was proposed by~\cite{mozgunov2017loss} which we use to construct the criterion for the allocation of patients. Criterion~(\ref{criterion}) still has the squared distance term in the nominator which ensures that it takes its minimum value $\delta(\cdot)=0$ at $p=\gamma$. At the same time, the denominator represents the variance of the probability of a binary event. Then, criterion~(\ref{criterion}) can be considered a score statistic which takes into account the uncertainty of the estimation object. In the context of the Phase I trial, the denominator can have one more interpretation. If $p=0$ or $p=1$ then $\delta(\cdot)=\infty$ meaning that patients would be never allocated to doses corresponding to $0$ or $1$ DLT probabilities. Indeed, the property of assigning of infinite values to the extreme values ``drives away'' the selection from the bounds to the neighbourhood of the interval of interest~$\gamma$ \citep{mozgunov2018,sainthilary2018}. Importantly, the criterion~(\ref{criterion}) also has an information-theoretic justification as it maximises the information gain in the trial with the area of the special interest, the neighbourhood of the maximum acceptable toxicity~\citep{mozgunov2017}. 

Applying the criterion to the illustration example above helps to address the uncertainty issue as $$\delta(\hat{p}_1=0.2,\gamma=0.3) = 1/16 \ {\rm and} \ \delta(\hat{p}_2=0.4,\gamma=0.3) = 1/24.$$ This means that $d_2$ should be selected for a next patient as follows from Inequality~(\ref{probabilities}). Note that a single point estimate of the criterion~(\ref{criterion}) already summarises the information about uncertainty in itself which can provide a potential computational benefits.

The target toxicity $\gamma$ is always less than $0.5$ in Phase I clinical trials. Consequently, if one would consider two point estimates which stand on the same squared distance $(\gamma-\theta)^2$ from the $\gamma$ (for $\theta < \gamma$), the criterion~(\ref{criterion}) favours a higher probability estimate due to the variance term in the denominator which is maximised at the point $p=0.5$. Indeed, the same rate of terms $p$ and $(1-p)$ in the denominator implies that overly toxic and overly safe doses are equally penalised. This, however, might contradict with the ethical issues of a trial.

\begin{figure}[h!]
  \centering
    \includegraphics[width=0.8\textwidth]{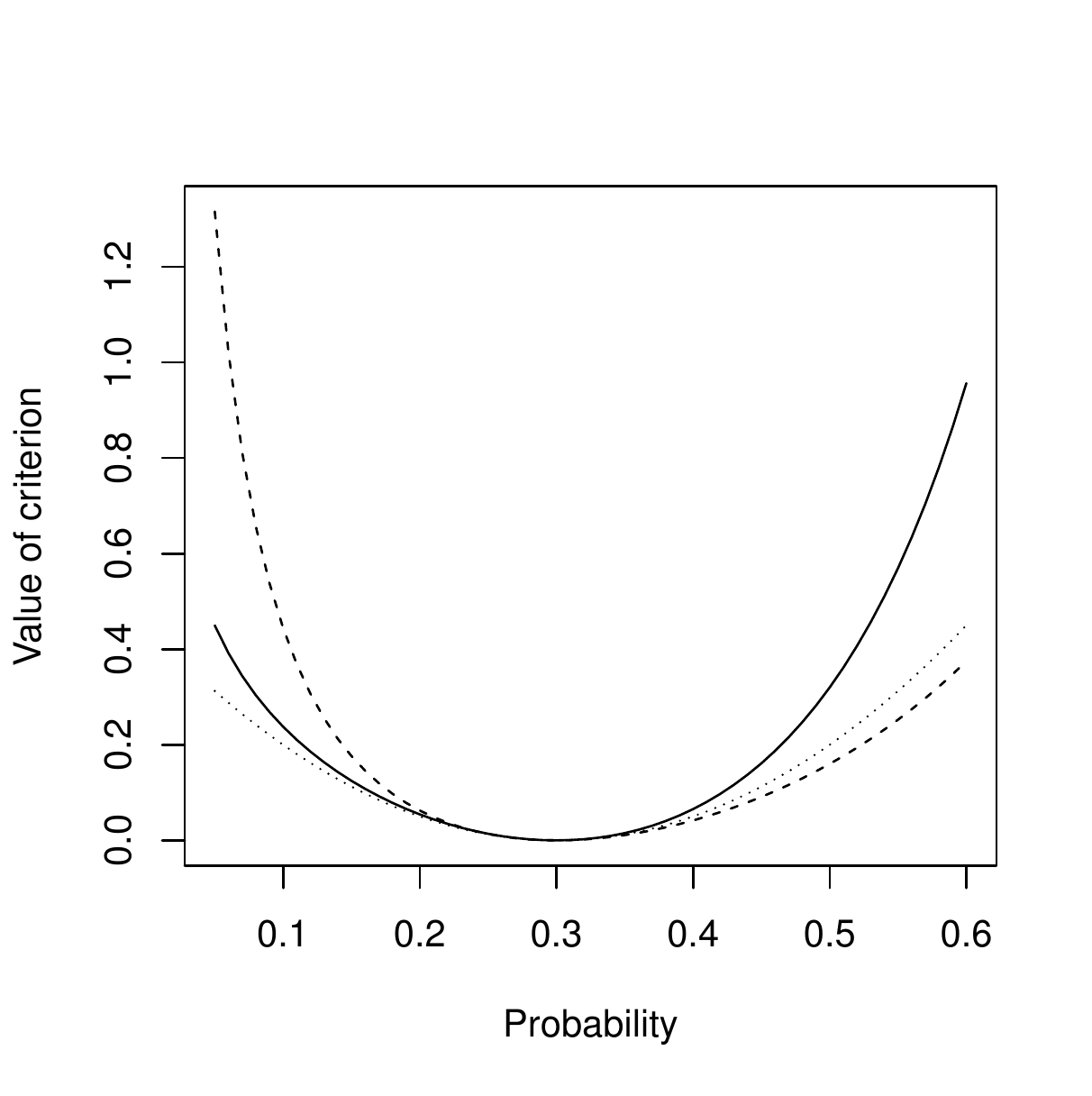}
    \caption{Squared distance criterion (dotted line) and CIBP criterion using the asymmetry parameter $a=1$ (dashed line), $a=0.5$ (solid line) for the target toxicity $\gamma=0.3$ and for different values~$p\in(0.05,0.6)$.}
    \label{fig:crt}
\end{figure}

To address this issue we generalise the criterion~(\ref{criterion}) to the case of asymmetric penalisation by including the asymmetry parameter $a$:
\begin{equation}
\delta(p,\gamma) = \frac{(p-\gamma)^2}{p^a(1-p)^{2-a}}.
\label{criterion1}
\end{equation}
The parameter $0<a<2$ corresponds to the penalisation of overly toxic doses and $2-a$ to overly safe doses. The constant $2$ is chosen to preserve the same rate of $p$ in both nominator and denominator to guarantee that $\delta \to 0$ when $p \to \gamma$ for all values of $\gamma$.  Clearly, values $0<a<1$ imply a more severe penalty for the allocation of patients to more toxic doses than to less toxic ones. Applying the proposed criterion with asymmetry parameter $a=0.5$ one can obtain that $$\delta(\hat{p}_1=0.2,\gamma=0.3,a=0.5) < \delta(\hat{p}_2=0.4,\gamma=0.3,a=0.5)$$ which means that dose $d_1$ would be selected due to the penalty on overly toxic doses. We would refer to the proposed criterion~(\ref{criterion1}) as to the \textit{Convex Infinite Bounds Penalization} (CIBP). An illustration of the squared distance criterion~(\ref{sq_distance}) and of the CIBP criterion~(\ref{criterion}) using $a=1$ and $a=0.5$ is given in Figure~\ref{fig:crt}.

The CIBP criterion (for both $a=1$ and $a=0.5$) goes to infinity faster than the squared distance as the probability $p$ approaches the lower bound. At the same time, for $a=1$ overly toxic doses are penalised less than by alternatives because corresponding values of the toxicity probability are located far from another boundary value $1$. The asymmetric CIBP criterion with $a=0.5$ solves this issue and penalises overly toxic doses more severely than both the squared distance and the symmetric CIBP. Note that all criteria behave similarly in the neighbourhood of the target $\gamma$. Overall, one can see that the properties of the proposed criterion allow resolving the ethical concern by setting an appropriate value of the parameter $a$. Further guideline on the choice of $a$ is given in the following section. 

\subsection{Choice of the asymmetry  parameter}
Firstly, note that the denominator alone is maximised at the point $p=a/2$. Then,  if $\hat{p}$ is an estimator of $p$ (depending on the approach, for instance, MLE or the Bayesian optimal estimator) the ``plug-in'' estimator of the CIBP criterion
\begin{equation}
\delta(\hat{p},\gamma) = \frac{(\hat{p}-\gamma)^2}{\hat{p}^a(1-\hat{p})^{2-a}}
\label{plugin}
\end{equation}
using $a=2\gamma$ leads to the same allocation as a plug-in estimator of the squared distance~(\ref{sq_distance}). Then, values $a < 2\gamma$ imply a more conservative allocation of patients than an original design which uses the squared distance criterion.

\begin{figure}[h!]
  \centering
    \includegraphics[width=0.8\textwidth]{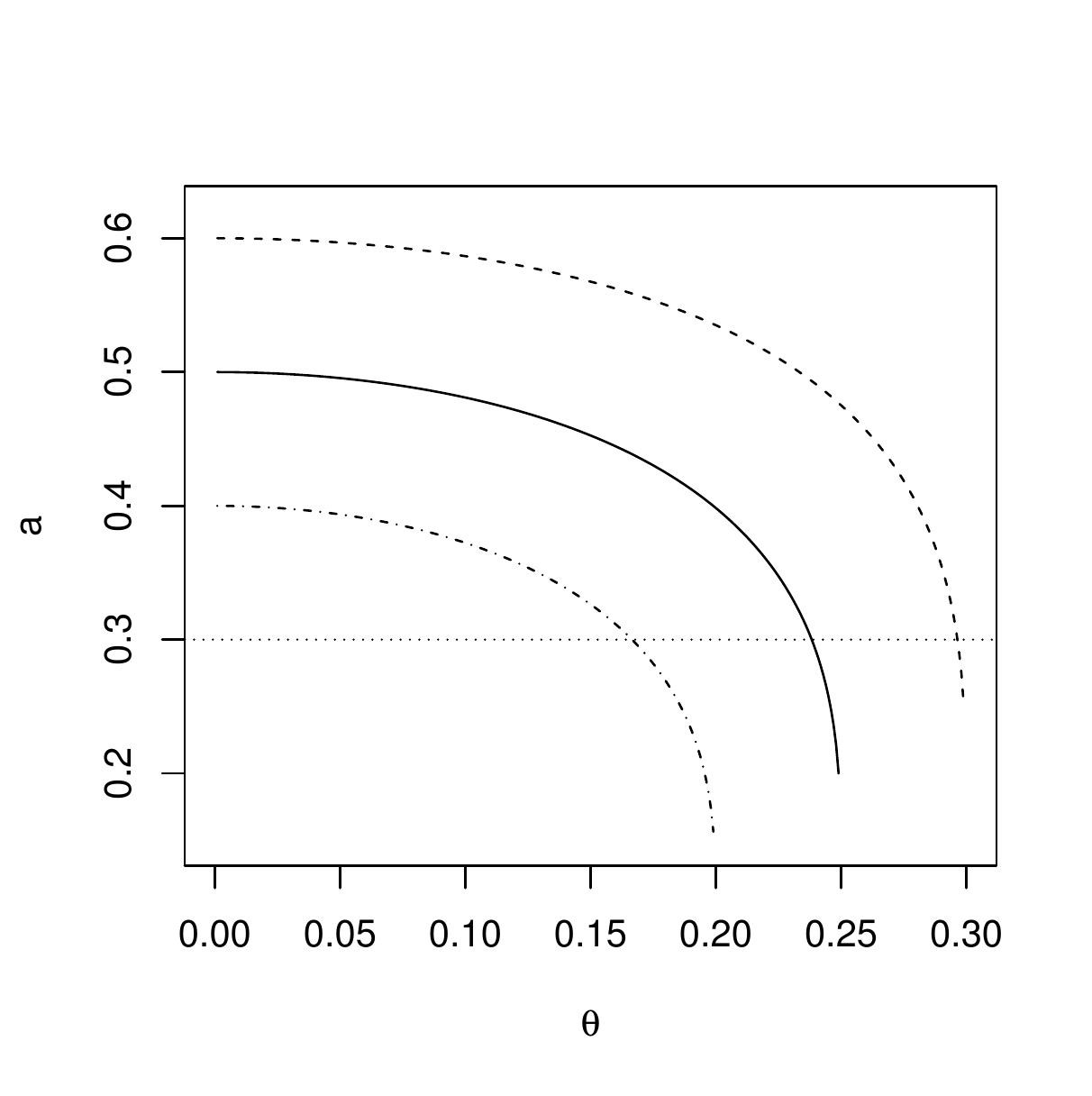}
    \caption{The values of parameter of asymmetry for $\gamma=0.20$ (dashed-dotted line) , $\gamma=0.25$ (solid line), $\gamma=0.30$ (dashed line) and different values of $\theta \in (0,0.35)$. The horizontal line corresponds to the choice of $a=0.30$ and corresponding half-width of intervals.}
    \label{fig:parameter}
\end{figure}

Secondly, the asymmetry parameter $a$ represents the trade-off between the ethical and uncertainty concerns. Then, for a sensible choice of $a$ we use the following condition. Consider an interval $(\gamma-\theta,\gamma+\theta)$. Assume that given two point toxicity probability estimates belonging to this interval and standing on the same squared distance from $\gamma$, one would like to select the lower toxicity estimate due to the safety concern. In other words, $(\gamma-\theta,\gamma+\theta)$ is the interval in which the safety issue is prioritised. Similarly, given two estimates lying outside of the interval $(\gamma-\theta,\gamma+\theta)$, but standing on the same squared distance, one would select that one which corresponds to a higher level of the uncertainty. Evidently, the estimates lying on the bounds of this interval should correspond to the same value of the CIBP criterion. Formally, solving

$$\frac{((\gamma-\theta)-\gamma)^2}{(\gamma-\theta)^a \left(1- (\gamma-\theta) \right)^{2-a}} =\frac{((\gamma+\theta)-\gamma)^2}{(\gamma+\theta)^a \left(1- (\gamma+\theta) \right)^{2-a}} $$
one can obtain that
$$ a = \frac{2}{1+A}$$ where
$$ A= \left( \log \frac{\gamma-\theta}{\gamma+\theta} \right) / \left( \log \frac{1-\gamma-\theta}{1-\gamma+\theta} \right)
$$
Then, for the fixed target value of $\gamma$ and the half-width of the interval~$\theta$, one can compute the corresponding value of $a$. Figure~\ref{fig:parameter} shows the dependence of the asymmetry parameter on the half-width $\theta$ and the target probabilities $\gamma=\{0.20,0.25,0.30\}$.

As $\theta \to 0$ (the uncertainty issue is prioritised), $a$ tends to $2\gamma$ which choice corresponds to the squared distance allocation rule as shown above. Increasing values of $\theta$ correspond to a wider interval in which an investigator prefers a lower toxicity estimate. Consequently, this corresponds to a more conservative allocation and to smaller values of $a$. Note that $a$ corresponding to $\theta \approx \gamma$ guarantees that for two estimates standing on the same squared distance from the target $\gamma$, the dose corresponding to the lower toxicity estimate would be always selected. For example, for the target value~$\gamma=0.25$ and the half-width $\theta=0.245$, the corresponding value of $a$ is close to $0.3$ (marked by the dotted horizontal line in Figure~\ref{fig:parameter}). 

In the next section, we recall the Bayesian continual reassessment method by~\cite{QPF90} and incorporate the proposed allocation criterion in the design.

\subsection{Bayesian continual reassessment method}
Consider a Phase I clinical trial with $m$ doses and $n$ patients. Assume that the DLT probability has the functional form
$$p_i=\psi(d_i,\beta)$$
where $\beta \in \mathbb{R}^h$ is a $h$-dimensional vector of parameters and $d_i,\ i=1,\ldots,m$ are standardised dose levels. Denote the prior distribution of $\beta$ by $f_0(.)$. Assume that $j$ patients have already been assigned to doses $d(1), \ldots, d(j)$ and binary responses $\mathbb{Y}_j=[y_1, \ldots, y_j]^{\rm T}$ were observed. The CRM updates the posterior distribution of $\beta$ using Bayes's Theorem
\begin{equation}
f_j(\beta)= \frac{f_{j-1}(\beta)\phi(d(j),y_j,\beta)}{\int_{\mathbb{R}^h} f_{j-1}(u)\phi(d(j),y_j,u) {\rm d} u}=\frac{f_0(\beta) \prod_{i=1}^{j}\phi(d(i),y_i,\beta)}{\int_{\mathbb{R}^h}  f_0(u) \prod_{i=1}^{j}\phi(d(i),y_i,u) {\rm d} u}
\label{posterior}
\end{equation}
where
$$\phi(d(j),y_j,\beta)=\psi(d(j),\beta)^{y_j}(1-\psi(d(j),\beta))^{1-y_j}.$$
Then, the posterior mean of the DLT probability for dose $d_i$ after $j$ patients is equal to
\begin{equation}
\hat{p}_i^{(j)}=\mathbb{E}(\psi(d_i,\beta)|\mathbb{Y}_j) = \int_{\mathbb{R}^h} \psi(d_i,u) f_j(u) { \rm d} u.
\label{estimate}
\end{equation}
As it was outlined above, the original design uses the following criterion. The dose $d_k$ minimising
\begin{equation}
(\hat{p}_i^{(j)} - \gamma)^2
\label{criterion_old}
\end{equation}
among all $d_1,\ldots, d_m$ is selected for the next group of patients. We propose to replace this step by the following allocation rule. 
The dose $d_k$  minimising
\begin{equation}
\mathbb{E} \left( \frac{\left(\psi(d_i,\beta)-\gamma\right)^2}{\psi(d_i,\beta)^a (1-\psi(d_i,\beta))^{2-a}}\right)
\label{criterion_new}
\end{equation}
among all $d_1,\ldots, d_m$ is selected for the next group of patients where the expectation is found w.r.t. to the posterior probability $f_j(\beta)$.  The procedure is repeated until the maximum number of patients, $n$, has been treated. As the uncertainty and the conservatism is important in the allocation only, we propose to use the squared distance~(\ref{criterion_old}) for the final MTD selection.

Many implementations of the CRM plug the mean value of $\beta$ in the model $\psi(d_i,\hat{\beta})$ instead of using the mean value, $\mathbb{E}(\psi(d_i,\beta)|\mathbb{Y}_j)$. While no noticeable difference is found in these approaches if a one-parameter model is used~\citep{iasonos2016}, it might affect the results significantly if more complex functions are considered. Therefore, the posterior mean of the new criterion is used. For consistency across all designs, we would also use the mean probability estimate while performing the original CRM design. 

We concentrate on the one-parameter power model 
\begin{equation}
 \psi(d_i,\beta) = d_i^{\exp (\beta)}
 \label{1parametermodel}
 \end{equation}
which was shown to be a powerful tool to identify the MTD. As the final remark, there are no concerns about the CRM design to be not aggressive enough. Therefore, we would concentrate on values $a \leq 2 \gamma$ in the rest of the work.

\section{Application to an actual clinical trial}
\subsection{Setting}
To illustrate the impact of the proposed allocation criterion, we revisit the results of the actual clinical trial of Everolimus in Patients With HER2-overexpressing Metastatic Breast Cancer (NCT00426556). The study considers $3$ regimens of Everolimus given together  with Paclitaxel and Trastuzumab~(PT):
\begin{enumerate}
\item Daily dosing of Everolimus 5mg plus PT ($d_1$)
\item Daily dosing of Everolimus 10mg plus PT ($d_2$)
\item Weekly dosing of Everolimus 30mg plus PT ($d_3$)
\end{enumerate}
The goal is to find the regimen corresponding to the target toxicity $\gamma=0.3$.
Note that the amount of the complimentary drugs is fixed during the trial and a clinician is confident in the monotonic relationship of toxicity probabilities for $d_1,\ldots,d_3$. Thus, the trial can be analysed using the tools for the single-agent trials. The aggregated data available by the end of the trial is given in Table~\ref{tab:illustration}. We revisit the results of this trial using the novel allocation criterion.

\begin{table}[ht]
\caption{\label{tab:illustration} \small{Aggregated data of the Everolimus trial }}
\footnotesize
\centering
\begin{tabular}{cccccccc}
  \hline
 Dose & $d_1$ & $d_2$ & $d_3$ \\
  \hline
   Number of Patients assigned & 6 & 17 & 10 \\
  Number of DLTs  & 3 & 6 & 7 \\ 
  \hline
\end{tabular}
\end{table}

We apply the CRM design using the one-parameter power model~\ref{1parametermodel} using the robust operational prior distribution $\beta \sim \mathcal{N} \left( 0,1.34\right)$~\citep{o1996likelihood,cheung2011} and the skeleton $(0.20,0.30,0.40)$ with an adequate spacing \citep{zohar2010} and implying that the prior MTD is $d_2$. We restrict the design so that the dose skipping is not allowed and enforce starting from the lowest dose. Patients are enrolled in cohorts of $3$. Note that the parameters of the design are the same for both the original CRM and the CRM using the novel allocation rule. The only difference is the criterion for the selection of doses. The original CRM uses the squared distance~(\ref{criterion_old}) while the CIBP design uses the criterion~(\ref{criterion_new}). Following the interpretation of the asymmetry parameter, we fix $a=0.3$ to favour less toxic selections in a wide interval of toxicity probabilities. The designs are implemented using the interactive functions of the \texttt{bcrm}-package by~\cite{BCRM}. We use the aggregated data from Table~\ref{tab:illustration} to generate the responses in one realisation of the trial. Clearly, DLTs indicated in Table~\ref{tab:illustration} can appear in any sequence. Therefore, we generate a random sample (without replacement) for each dose to have a specific order of DLTs. We fix this order for both trials. The only exception is that the realisation for the first cohort is chosen by us. We consider the influence of this choice later.

\subsection{Illustration}
\begin{figure}[h!]
  \centering
    \includegraphics[width=1\textwidth]{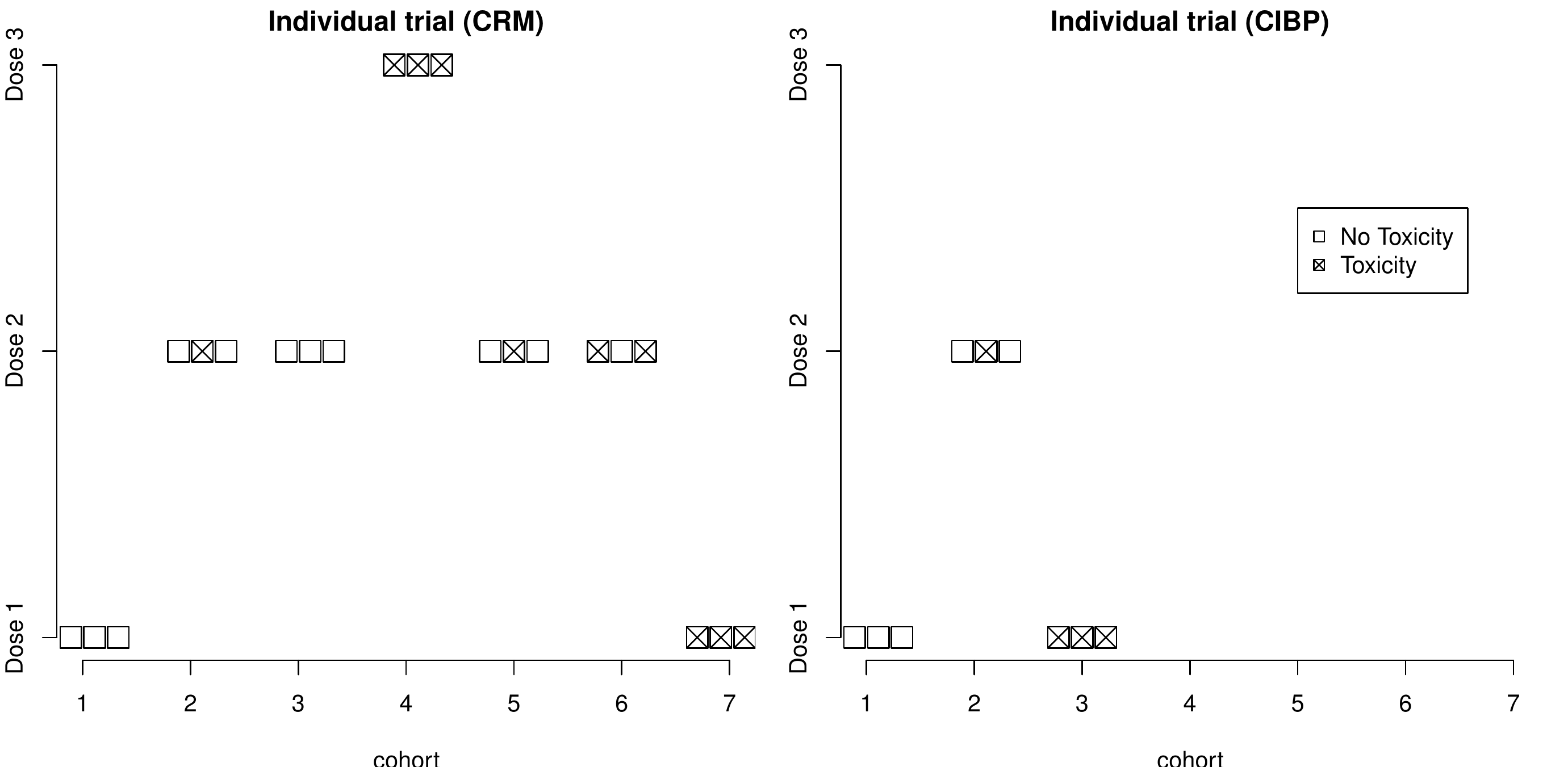}
    \caption{Allocation of $7$ cohorts in the individual Everolimus trial.}
    \label{fig:illustration}
\end{figure}

The first $3$ patients are assigned to $d_1$ by construction. We begin by assuming that all $3$ patients have not experienced DLTs. The sequential dose selections for the CRM and CIBP designs, in this case, are given in Figure~\ref{fig:illustration}. The values of the criteria after each cohort are given in Table~\ref{tab:ill_result}.

\begin{table}[ht]
\caption{\label{tab:ill_result} \small{The values of criterion (\ref{criterion_old}) used by the CRM and criterion (\ref{criterion_new}) used by CIBP in the individual trial after each cohort. The value of criterion corresponding to the dose selected for next cohort by each design is in \textbf{bold}}}
\footnotesize
\centering
\begin{tabular}{c|ccc|cccc}
  \hline
  &         \multicolumn{3}{@{}c}{{CRM}}  & \multicolumn{3}{@{}c}{{CIBP}}\\
  \hline
  Cohort 1 & 0.031  & \textbf{0.012}  & 0.002  & 0.62 & \textbf{0.47} & 0.45  \\
 Cohort 2  & 0.01 & \textbf{0.001}  & 0.005  &  \textbf{0.10} & 0.12 & 0.21 \\ 
  Cohort 3  & 0.03 & 0.013  & \textbf{0.001} &  \textbf{0.30} & 0.67 & 1.41 \\ 
 Cohort 4  & 0.003 & \textbf{0.002} & 0.003 &    &  \\ 
 Cohort 5  & 0.003 & \textbf{0.002}  & 0.024 &    &  \\ 
 Cohort 6  & \textbf{0.000}  & 0.009  & 0.039 &    &  \\ 
 Cohort 7  & \textbf{0.014} &  0.048 &0.096 &    &  \\ 
 \hline
\end{tabular}
\end{table}

After no DLTs were observed for the first cohort, CRM and CIBP allocate the second cohort of patients to $d_2$ for which one patient experiences a DLT. Given this toxic outcome, CRM recommends to stay at $d_2$ for the third cohort. In contrast, CIBP recommends returning to the previous dose level due to the conservatism of the criterion. Then, after all patients in cohort 3 (using CIBP) experienced the DLT, the trial would be terminated by a clinician due to safety. At the same time, the trial using the original CRM design proceeds. After no DLTs were observed for cohort 3, $d_3$ is recommended for cohort 4 in which all patients have DLT. This leads to de-escalation to $d_2$ and after 2 cohorts for which 3 patients (out of 6) had DLT and further de-escalation to $d_1$. All 3 patients in cohort 7 experienced DLTs and a clinician terminates a trial due to toxicity. Overall, while the CRM assigned 21 patients and $10$ of them experienced DLTs to come to the same conclusion as CIBP, the novel criterion allows to reduce the sample size to $9$ patients with $4$ toxicity outcomes only. 

The illustration above demonstrates the allocation if no toxicity outcomes are observed in cohort 1, but other possibilities should be considered as well based on aggregated data. Clearly, the other possibilities are 1, 2 and 3 DLTs in the first cohort. Considering these scenarios, it was found that both designs lead to the same allocation of patients and never escalate from dose $d_1$.  It follows that the novel allocation rule leads to the same MTD selection in all possible sequence of outcomes, but results in fewer or similar number of toxic responses. This motivates a further investigation of the novel criterion in a comprehensive simulation study.

\section{Comparison to the original CRM \label{sec:crm}}
\subsection{Setting}
\begin{figure}[h!]
  \centering
    \includegraphics[width=1\textwidth]{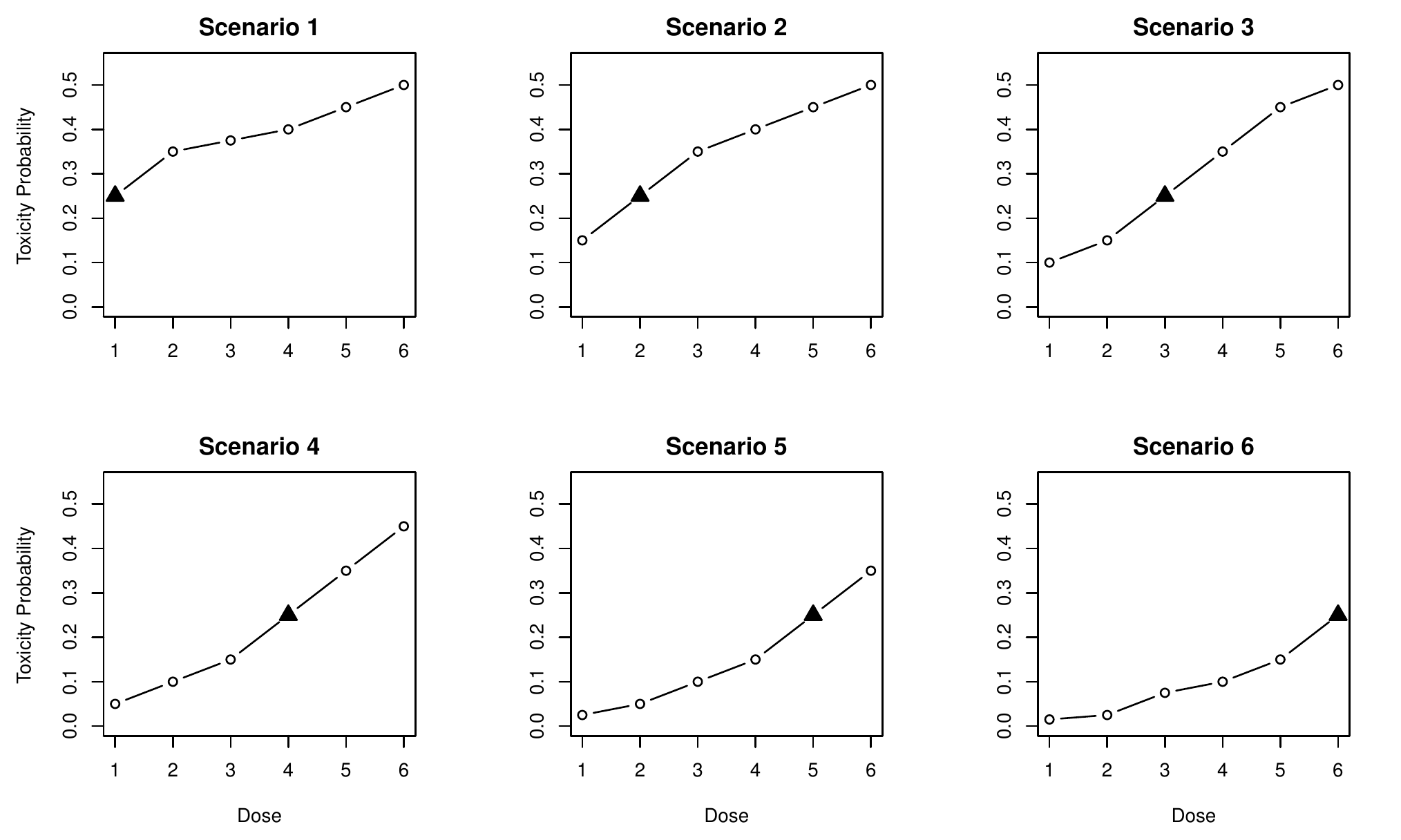}
    \caption{Six considered dose-toxicity scenarios for the comparison to the CRM. The MTD is marked by the black triangle.}
    \label{fig:scenarios}
\end{figure}

In this section, we compare the performance of the proposed criterion against the squared distance criterion both applied to the one-parameter power model. The single-agent Phase I trial with $m=6$ doses and $n=30$ patients is considered. The goal is to find the MTD corresponding to $\gamma=0.25$. We consider $6$ dose-toxicity scenarios with the target doses located at the dose corresponding to scenario's number. The shapes of the dose-toxicity are given in Figure~\ref{fig:scenarios}.
Toxicity scenarios were chosen ``equally difficult'' in terms of the optimal non-parametric benchmark \citep{benchmark}. It allows comparing the proportion of correct selections (PCS) between different scenarios. We specify the skeleton for the one-parameter power model using the package \texttt{dfcrm} and the function \texttt{getprior} using that the prior MTD is $d_2$ and the half-width of the equivalent interval is $0.05$. The prior distribution of the parameter is chosen to be $\beta \sim \mathcal{N} \left(0,1.34\right)$ \citep{o1996likelihood,cheung2011,cheung2013}. Different skeletons corresponding to $d_3$ and $d_4$ being the MTD are also investigated and the corresponding (quantitatively similar) results are given in the Appendix. Again, both the CRM with the standard allocation criterion and the CRM with the novel criterion use the same model parameters and the only difference is the selection rule. This allows to link the observed differences to the criteria choice only. We study (i) the proportion of the correct selections (PCS) and (ii) the proportion of patients experienced a toxic response. We consider different values of $a=\{0.3,0.4,0.5\}$ corresponding to the approximate half-width of the intervals $\theta\approx\{0.00,0.20,0.25\}$. 

We denote the CRM with the new escalation criterion using parameter $a$ by CIBP(a).  The characteristics of all the models compared are evaluated in \texttt{R} \citep{Rcore} using the \texttt{bcrm}-package by \cite{BCRM}. To accommodate the new criterion, the corresponding modifications to the package were made. 

\subsection{Operating characteristics}
Proportions of each dose selections and proportions of patients experienced a DLT for  CRM and for CIBP are given in Table~\ref{tab:result1}. We use $40000$ simulations to declare any difference above $1\%$ as a significant one.

\begin{table}[h!]
\caption{\label{tab:result1} \small{Proportions of each dose selections and proportions of DLTs (Tox) in one trial for CRM and CIBP using $a=\{0.3,0.4,0.5\}$. Results are based on $40000$ simulations.}}
\footnotesize
\centering
\begin{tabular}{cccccccc}
  \hline
 & $d_1$ & $d_2$ & $d_3$ & $d_4$ & $d_5$ & $d_6$ & Tox \\ 
  \hline
        \multicolumn{7}{@{}c}{{Scenario 1}}\\
Toxicity & \textbf{25.00} & 35.00 & 37.50 & 40.00 & 45.00 & 50.00  \\ 
  CIBP(0.3) & 69.18 & 21.65 & 6.18 & 2.27 & 0.61 & 0.11 & 28.31  \\ 
   CIBP(0.4) & 66.40 & 22.20 & 7.25 & 3.08 & 0.91 & 0.16 & 29.46  \\ 
      CIBP(0.5)& 64.12 & 22.25 & 8.49 & 3.80 & 1.15 & 0.18 & 30.58 \\  
 CRM  & 65.59 & 21.16 & 8.22 & 3.79 & 1.07 & 0.17 & 30.17 \\ 
   \multicolumn{7}{@{}c}{{Scenario 2}}\\
   Toxicity & 15.00 & \textbf{25.00} & 35.00 & 40.00 & 45.00 & 50.00  \\ 

  CIBP(0.3) & 24.06 & 47.88 & 21.98 & 5.00 & 0.93 & 0.15 & 23.33  \\ 
CIBP(0.4) & 24.00 & 46.95 & 22.28 & 5.39 & 1.19 & 0.19 & 24.87 \\ 
CIBP(0.5) & 23.97 & 46.12 & 22.20 & 6.02 & 1.46 & 0.24 & 26.45  \\      
CRM  & 25.41 & 45.76 & 21.36 & 5.96 & 1.27 & 0.24 & 26.10 \\ 
          \multicolumn{7}{@{}c}{{Scenario 3}}\\
         Toxicity & 10.00 & 15.00 & \textbf{25.00} & 35.00 & 45.00 & 50.00 \\ 

  CIBP(0.3) & 4.26 & 25.61 & 46.48 & 20.20 & 3.16 & 0.28 & 20.91\\ 
 CIBP(0.4) & 4.08 & 25.53 & 46.25 & 20.57 & 3.23 & 0.34 & 22.56  \\ 
 CIBP(0.5) & 3.77 & 25.64 & 46.49 & 20.46 & 3.31 & 0.32 & 24.21  \\ 
    CRM  & 3.91 & 26.66 & 45.62 & 20.37 & 3.06 & 0.37 & 23.97 \\ 
  \multicolumn{7}{@{}c}{{Scenario 4}}\\
  Toxicity & 5.00 & 10.00 & 15.00 & \textbf{25.00} & 35.00 & 45.00 &  \\ 
 
  CIBP(0.3) & 0.22 & 5.01 & 27.27 & 44.62 & 19.65 & 3.23 & 19.36 \\ 
 CIBP(0.4) & 0.17 & 4.78 & 26.64 & 45.66 & 19.59 & 3.15 & 20.99 \\ 
 CIBP(0.5)& 0.18 & 4.74 & 27.16 & 45.74 & 19.36 & 2.83 & 22.43  \\ 
 CRM  & 0.18 & 4.50 & 27.82 & 45.32 & 19.15 & 3.03 & 22.43  \\ 
    \multicolumn{7}{@{}c}{{Scenario 5}}\\
     Toxicity & 2.50 & 5.00 & 10.00 & 15.00 & \textbf{25.00} & 35.00 \\ 

  CIBP(0.3) & 0.00 & 0.34 & 6.54 & 27.67 & 43.34 & 22.11 & 17.71  \\ 
CIBP(0.4) & 0.00 & 0.31 & 5.89 & 27.77 & 44.12 & 21.89 & 19.24  \\ 
CIBP(0.5) & 0.00 & 0.33 & 5.50 & 28.06 & 44.84 & 21.28 & 20.73\\ 
    CRM  & 0.01 & 0.27 & 5.46 & 28.89 & 44.10 & 21.28 & 20.56  \\ 
  \multicolumn{7}{@{}c}{{Scenario 6}}\\
 Toxicity & 1.50 & 2.50 & 7.50 & 10.00 & 15.00 & \textbf{25.00}  \\ 
  CIBP(0.3) & 0.00 & 0.04 & 2.97 & 10.42 & 26.84 & 59.72 & 15.25\\ 
   CIBP(0.4) & 0.00 & 0.05 & 2.30 & 9.55 & 27.53 & 60.58 & 16.53  \\  
    CIBP(0.5)  & 0.00 & 0.04 & 1.68 & 7.31 & 27.71 & 63.26 & 17.98  \\  
 CRM  & 0.00 & 0.05 & 1.88 & 8.65 & 28.89 & 60.53 & 17.34  \\
\end{tabular}
\end{table}

\noindent Comparing the performance of CIBP for different values of the asymmetry parameter, one can see that more conservative allocation and selection correspond to CIBP(0.3). The greatest difference can be seen in scenarios 1 and 6. The increase in $a$ from $0.3$ to $0.5$ leads to an increase in the PCS by 5\% in the toxic scenario 1 and to a decrease in the PCS by 3.5\% in the flat scenario 6. The differences in the rest of scenarios are smaller, but still significant.
Overall, greater values of $a$ favour higher doses to be selected and lead to a higher mean proportion of toxic responses with the difference around 2-3\% between CIBP(0.3) and CIBP(0.5) in all scenarios.

Regarding the comparison of CIBP and CRM, one can find that CIBP(0.4) has a similar PCS, but also a smaller proportion of toxic responses in all considered scenarios. The CIBP(0.5) performs similar (scenarios 2-5) or better (scenario 6) than CRM at the cost of 1\% decrease in the PCS in scenario 1. The most noticeable difference can be observed by comparing CRM to CIBP(0.3). In terms of the PCS, CIBP(0.3) outperforms the CRM by 4\% and 2\% in the most toxic scenarios 1, 2 and shows the comparable performance in rest of scenarios. At the same time, CIBP(0.3) outperforms the CRM in terms of the proportion of toxic responses in all considered scenario by nearly 3\% in scenarios 2-5 and by 2\% in scenarios 1 and 6. While the margin of the difference might be seen to be negligibly small, this improvement results in nearly 1 fewer patient experienced a DLT. Therefore, this design is more ethical as it exposes fewer patients to more toxic doses while leading to no changes or an increase in the PCS.

Another valuable feature of the novel allocation criterion is an additional flexibility which allows controlling the number of toxic responses directly. A clinician can choose the parameter $a$ based on their conservatism and on the range of scenarios of interest. For instance, a clinician might be ready to sacrifice the PCS in the flat scenario 6 for the sake of not selecting overly toxic dose in scenario 1. The new criterion enables such an option. At the same time, the design preserves its simplicity and does not result in any extra computational costs.

\section{Comparison to the EWOC and its modifications}
\subsection{Setting}
Alternative criteria for solving the ethical and uncertainty issues using the two-parameter logistic model
$$\psi(d_i,\beta_1,\beta_2) = \frac{\exp (\beta_1+\beta_2d_i)}{1+\exp (\beta_1+\beta_2d_i)} $$
were proposed by \cite{ewoc} using the EWOC design. However, as stated above the EWOC can result in a systematic underestimation of the MTD. Therefore, some modifications were proposed by \cite{mourad2010,wheeler2017}. The main idea beyond the modifications is to use a changing parameter $\alpha_n$ in the criterion~(\ref{ewoc_distance}) rather than a fixed value $\alpha$. The detailed description of these modifications can be found in \cite{wheeler2017}. Alternatively, the BLRM method by \cite{neuenschwander2008}also using the two-parameter logistic model and a loss function, can be used. In this section, we compare the performance of the one-parameter CRM design using the novel allocation rule to these designs.

\begin{figure}[h!]
  \centering
    \includegraphics[width=1\textwidth]{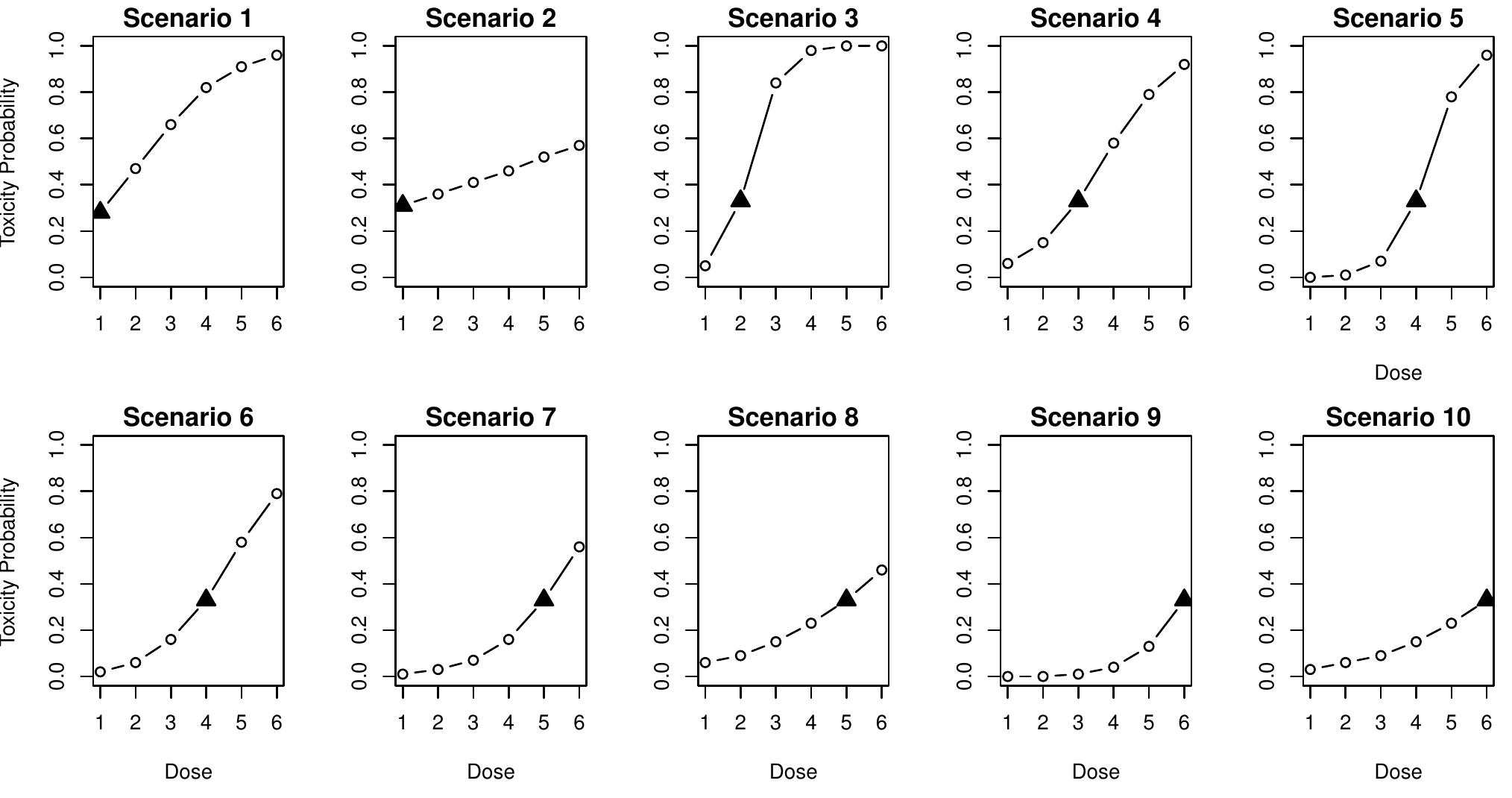}
    \caption{Ten considered dose-toxicity scenarios for the comparison to EWOC. The MTD is marked by the black triangle.}
    \label{fig:scenarios2}
\end{figure}

We consider the setting by \cite{wheeler2017} for discrete dose levels. There are $n=40$ patients and $m=6$ doses in the trial. The goal is to find the MTD corresponding to the $\gamma=0.33$.  The original scenarios are given in Figure~\ref{fig:scenarios2}.

The prior distribution of $\beta$ is specified as in the previous section. The only difference is the skeleton which is now set using the same information as by \cite{wheeler2017}: the prior MTD is $d_3$. Assuming that ethical issues are of the greater interest in this trial we consider $a=\{0.5,0.25,0.10\}$.

We compare the performance of the proposed approach to
\begin{itemize}
\item EWOC design - the original EWOC design using fixed $\alpha=0.25$
\item TR design by~\citep{mourad2010} using $\alpha_2 = \ldots=\alpha_9=0.25$, $ \alpha_n=\min \left(\alpha_{n-1}+0.05,0.50 \right)$ for all future patients.
\item Toxicity-dependent feasibility bound design (TDFB) by~\cite{wheeler2017} using
$$ \alpha_{n+1} = \min \left( 0.50, \alpha_{min} + (0.50-\alpha_{min} \frac{n-1-\sum_{i=1}^n y_i}{S} \right) $$
where $n-1-\sum_{i=1}^n y_i$ is number of patients with no DLTS, $\alpha_{min}=0.25$ and $S=12\frac{2}{3}$. For both modifications of the EWOC design above we use the parameters as in \cite{wheeler2017}.
\item Design by \cite{neuenschwander2008} (BLRM) which uses the loss function for the decision. Following the original proposal we use the same bivariate normal prior distribution for parameters as in the original work and adapt the toxicity intervals for the loss function for $\gamma=0.33$ 
$$L=
\begin{cases}
1 \ {\rm if} \ p \in (0.00,0.26) \\
0 \ {\rm if} \ p \in (0.26,0.41) \\
1 \ {\rm if} \ p \in (0.41,0.66) \\
2 \ {\rm if} \ p \in (0.66,1.00)
\end{cases}
$$
\end{itemize}

Following \cite{wheeler2017} we study the performance of designs in terms of (i) \textit{Accuracy} 
$$ \mathcal{A} =  1 - m \frac{\sum_{i=1}^m \left(p_i-\gamma\right)^2 \pi_i }{\sum_{i=1}^m \left(p_i-\gamma\right)^2}$$
where $p_i$ is the true toxicity probability for $d_i$ and $\pi_i$ is the probability to select $d_i$ and in terms of $(ii)$ mean number of toxic responses (DLTs). As many different  scenarios are considered, one can expect that one design would outperform another in some of them \citep{wages2015comment}. Therefore, we focus on average performance: the (geometric) mean accuracy and the mean number of DLTs across all scenarios.

\subsection{Operating characteristics}
\begin{figure}[h!]
  \centering
    \includegraphics[width=1\textwidth]{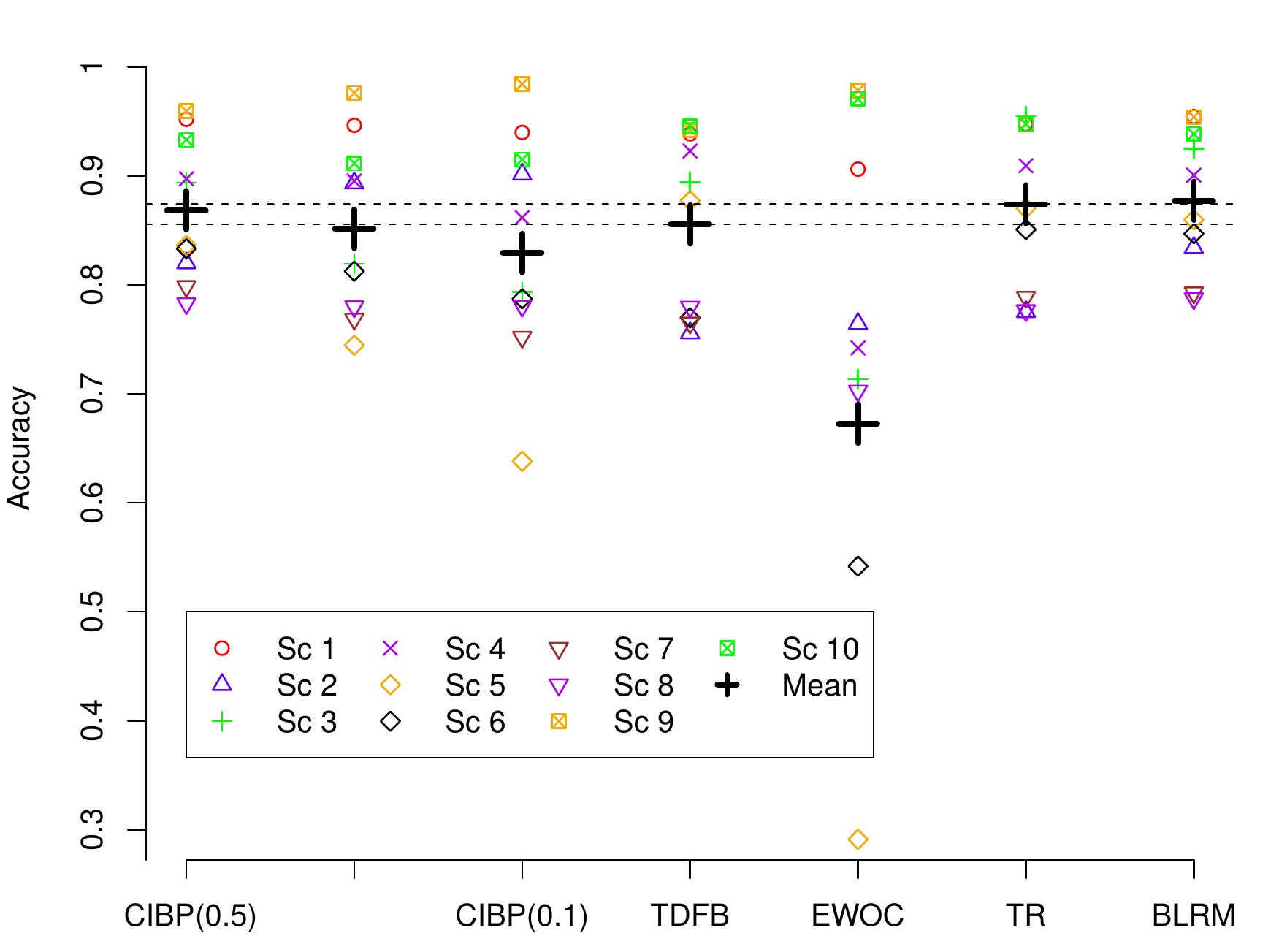}
    \caption{Accuracy indices and mean accuracy indices for CIBP using $a=\{0.5,0.25,0.10\}$, TDFB, EWOC, TR and BLRM designs. The upper dashed horizontal line corresponds to the accuracy of the TR and the lower one to the accuracy of the TDFB. Results are based on 2000 simulations.}
    \label{fig:accuracy}
\end{figure}

The accuracy index for the CIBP using $a=\{0.5,0.25,0.10\}$, TDFB, EWOC, TR and BLRM design are given in Figure~\ref{fig:accuracy}.
Comparing CIBP for different values of $a$, one can see that the mean accuracy decreases with parameter $a$. Due to a more conservative allocation, fewer patients are assigned to doses in the neighbourhood of the MTD. This results in a lower PCS and a lower accuracy index. The decrease in the accuracy index is, however, rather small - from $0.87$ using $a=0.5$ to $0.83$ using $a=0.1$. The most noticeable drop across scenarios can be found in scenario 5 - $0.20$.
Clearly, the variance of the accuracy indexes increases with decreasing $a$ - a more conservative design leads to a better performance in toxic scenarios 1-3 for the cost of a less accurate performance in flat scenarios 8-10.

\begin{figure}[h!]
  \centering
    \includegraphics[width=1\textwidth]{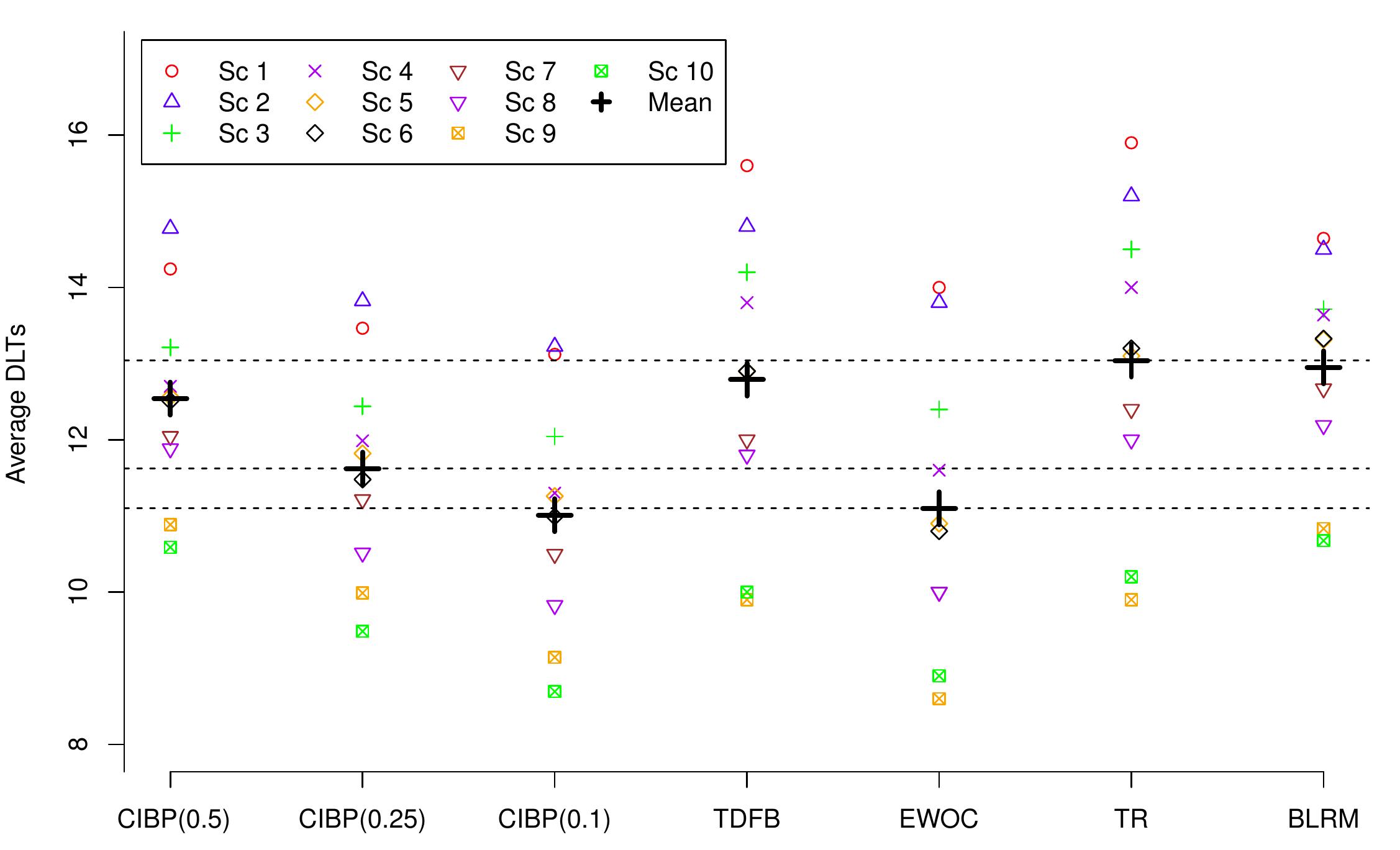}
    \caption{DLTs and mean number of DLTs for the CIBP using $a=\{0.5,0.25,0.10\}$, TDFB, EWOC, TR and BLRM designs. The upper, middle and lower dashed horizontal lines correspond to performance of TR, CIBP(0.25) and EWOC, respectively. Results are based on 2000 simulations.}
    \label{fig:dlts}
\end{figure}

Comparing different approaches, TR, BLRM and CIBP(0.5) correspond to the highest mean performance. However, TR corresponds to a slightly greater variability of the accuracy indices between scenarios. The TDFB design performs comparably to CIBP(0.25) design both in terms of the mean accuracy and the accuracy variability. As expected, the original EWOC results in the least mean accuracy index due to the large MTD underestimation in scenarios 5-7. The mean accuracy index associated with the most conservative CIBP(0.10) is greater than associate with EWOC by $0.15$. The accuracy indices themselves are of a limited interest when considered alone with taking into account the safety. The mean number of DLTs in all considered designs are given in Figure~\ref{fig:dlts}.

Regarding CIBP, while lower values of $a$ resulted in a lower accuracy they also result in fewer DLTs on average across all scenarios. As the result, the mean number of DLTs is decreased by approximately $2$ toxic responses comparing  CIBP(0.5) and  CIBP(0.10). Considering different designs, it is of interest to compare designs which have comparable mean accuracy indices. The TR and BLRM designs result in nearly $0.5$ more toxic responses on average than CIBP(0.5). The TDFB design that has the accuracy similar to CIBP(0.4) results in nearly $1.5$ excessive toxic responses than the novel design. Interestingly, the EWOC, which had the mean accuracy index lower than CIBP(0.1) by $0.15$, results in a similar mean number of DLTs across all scenarios as well as CIBP(0.10). Interestingly, it also leads to nearly one additional patient with a toxic response in highly toxic scenarios 1 and 2.

Overall, in contrast to TDFB and TR, the proposed design does not change the parameter of conservatism as more patients are trialled. It requires only one extra parameter to be specified. However, one can find a value of parameter $a$ that would lead to similar  accuracy index, but a fewer mean number of DLTs in all scenarios. At the same time, the safest version of the CIBP design results in comparable to the EWOC mean number of DLTs across all scenarios, but also in a noticeably greater accuracy.

\section{Discussion}
The novel dose-escalation criterion for the allocation of patients is introduced in this work. The criterion requires only one additional parameter, which has clear intuitive interpretation and can be easily tuned according to the purposes of the investigator, to be specified. A guideline on the choice of parameter is also given. We incorporate this criterion into the one-parameter power Bayesian CRM method which has shown to be a powerful tool to identify the MTD. It is found that the proposed design results in fewer number of toxic responses in a trial than the original CRM with no loss in the probability of correct selections. Comparing the novel design to alternative approaches, we have found that there exists a value of the asymmetry parameter such that the design would have a similar accuracy, but a lower mean number of toxic responses. Therefore, the new criterion allows to make model-based designs more ethical as it decreases the number of patients experienced DLTs but does not lead to any decrease in accuracy. 

Importantly, the novel criterion proposed in this work can be applied to any parametric model and not limited to the one-parameter model. 
The application of the criterion was demonstrated in the context of a single-agent trial only. As there are generalisations of the CRM design for more complex studies, it is also of interest to consider the application of the novel allocation rule to dose-combination and dose-schedule trials including the case of delayed toxicity responses.

\section*{Acknowledgement}

This project has received funding from the European Union’s
Horizon 2020 research and innovation programme under the 
Marie Sklodowska-Curie grant agreement No 633567 and by Prof Jaki's Senior Research Fellowship (NIHR-SRF-2015-08-001) 
supported by the National Institute for Health Research. The views expressed in this publication are those of the authors and not necessarily those of the NHS, the National Institute for Health Research or the Department of Health.

\bibliographystyle{rss}
\bibliography{mybib}  

\clearpage
\section*{Appendix}
\begin{table}[h!]
\caption{\label{tab:appendix1} \small{Proportions of doses selections and the proportion of DLTs (TR) in one trial for the CRM and the CIBP using $a=\{0.3,0.4,0.5\}$ and the prior parameter as described in Section \ref{sec:crm}, but the prior MTD $d_3$ is used for the skeleton construction. Results are based on $40000$ simulations.}}
\footnotesize
\centering
\begin{tabular}{ccccccccccc}
  \hline
 & $d_1$ & $d_2$ & $d_3$ & $d_4$ & $d_5$ & $d_6$ & TR  \\ 
  \hline
     &    \multicolumn{6}{@{}c}{{Scenario 1}}\\
Toxicity & \textbf{25.00} & 35.00 & 37.50 & 40.00 & 45.00 & 50.00 \\ 
  CIBP (0.3) & 67.59 & 22.50 & 6.44 & 2.60 & 0.74 & 0.13 & 28.73 \\ 

   CIBP (0.4) & 64.12 & 22.53 & 8.29 & 3.77 & 1.08 & 0.21 & 29.98 \\ 
     CIBP (0.5) & 61.78 & 22.23 & 9.61 & 4.80 & 1.30 & 0.28 & 31.31  \\ 

     CRM  & 63.27 & 21.28 & 9.45 & 4.52 & 1.28 & 0.21 & 31.02  \\

     &    \multicolumn{6}{@{}c}{{Scenario 2}}\\
   Toxicity & 15.00 & \textbf{25.00} & 35.00 & 40.00 & 45.00 & 50.00   \\ 
  CIBP (0.3) & 22.86 & 47.00 & 23.37 & 5.45 & 1.15 & 0.17 & 23.94\\ 
 
  CIBP (0.4) & 22.89 & 45.59 & 23.64 & 6.25 & 1.40 & 0.23 & 25.64  \\ 
    CIBP (0.5) & 23.47 & 44.65 & 23.12 & 6.97 & 1.49 & 0.29 & 27.14  \\ 

  CRM  & 24.38 & 44.26 & 22.64 & 6.81 & 1.64 & 0.27 & 27.23 \\ 

     &    \multicolumn{6}{@{}c}{{Scenario 3}}\\
         Toxicity & 10.00 & 15.00 & \textbf{25.00} & 35.00 & 45.00 & 50.00 \\ 
  CIBP (0.3) & 3.76 & 23.83 & 46.95 & 21.62 & 3.51 & 0.33 & 21.74\\ 
 
  CIBP (0.4) & 3.84 & 23.95 & 46.45 & 21.98 & 3.47 & 0.32 & 23.25\\ 
    CIBP (0.5) & 3.53 & 23.99 & 46.36 & 22.17 & 3.56 & 0.40 & 25.09  \\ 

  CRM  & 3.52 & 24.57 & 46.51 & 21.54 & 3.50 & 0.37 & 25.17  \\ 

     &    \multicolumn{6}{@{}c}{{Scenario 4}}\\
  Toxicity & 5.00 & 10.00 & 15.00 & \textbf{25.00} & 35.00 & 45.00  \\ 
  CIBP (0.3) & 0.17 & 4.45 & 25.60 & 46.06 & 20.42 & 3.29 & 20.10 \\ 
 
  CIBP (0.4) & 0.16 & 4.15 & 25.51 & 46.38 & 20.60 & 3.20 & 21.62  \\ 
    CIBP (0.5) & 0.15 & 4.00 & 25.66 & 46.66 & 20.38 & 3.14 & 23.37 \\ 

  CRM & 0.18 & 4.08 & 26.25 & 46.39 & 20.24 & 2.85 & 23.40\\ 

     &    \multicolumn{6}{@{}c}{{Scenario 5}}\\
     Toxicity & 2.50 & 5.00 & 10.00 & 15.00 & \textbf{25.00} & 35.00  \\ 
  CIBP (0.3) & 0.00 & 0.27 & 5.71 & 27.24 & 44.13 & 22.64 & 18.39  \\ 
  CIBP (0.4) & 0.00 & 0.24 & 5.06 & 26.41 & 45.88 & 22.41 & 19.96 \\ 
   CIBP (0.5) & 0.00 & 0.27 & 4.97 & 27.23 & 45.48 & 22.05 & 21.41 \\ 

    CRM  & 0.00 & 0.25 & 4.58 & 27.98 & 45.88 & 21.31 & 21.56 \\

     &    \multicolumn{6}{@{}c}{{Scenario 6}}\\
 Toxicity & 1.50 & 2.50 & 10.00 & 10.00 & 15.00 & \textbf{25.00}   \\ 
   CIBP (0.3) & 0.00 & 0.02 & 2.42 & 9.37 & 26.83 & 61.36 & 15.83 \\ 
 
  CIBP (0.4) & 0.00 & 0.02 & 1.74 & 7.62 & 27.18 & 63.44 & 17.20 \\ 
    CIBP (0.5) & 0.00 & 0.05 & 1.74 & 7.75 & 27.89 & 62.58 & 18.20 \\ 

  CRM  & 0.00 & 0.04 & 1.29 & 6.63 & 28.55 & 63.49 & 18.36 \\

\end{tabular}
\end{table}

\begin{table}[ht]
\caption{\label{tab:appendix2} \small{Proportions of doses selections and the proportion of DLTs (TR) in one trial for the CRM and the CIBP using $a=\{0.3,0.4,0.5\}$ and the prior parameter as described in Section \ref{sec:crm}, but the prior MTD $d_4$ is used for the skeleton construction. Results are based on $40000$ simulations.}}
\footnotesize
\centering
\begin{tabular}{ccccccccccc}
  \hline
 & $d_1$ & $d_2$ & $d_3$ & $d_4$ & $d_5$ & $d_6$ & TR \\ 
  \hline
     &    \multicolumn{6}{@{}c}{{Scenario 1}}\\
Toxicity & \textbf{25.00} & 35.00 & 37.50 & 40.00 & 45.00 & 50.00   \\ 
   CIBP  (0.3) & 64.38 & 23.48 & 7.74 & 3.21 & 1.03 & 0.15 & 29.25  \\ 

  CIBP  (0.4) & 62.80 & 22.43 & 9.04 & 4.15 & 1.35 & 0.24 & 30.56 \\ 
   CIBP  (0.5) & 59.16 & 22.61 & 10.51 & 5.68 & 1.73 & 0.31 & 31.93 \\

  CRM  & 60.09 & 21.50 & 10.73 & 5.63 & 1.74 & 0.30 & 31.96  \\ 

     &    \multicolumn{6}{@{}c}{{Scenario 2}}\\
   Toxicity & 15.00 & \textbf{25.00} & 35.00 & 40.00 & 45.00 & 50.00  \\ 
      CIBP  (0.3) & 21.23 & 46.23 & 24.93 & 6.07 & 1.36 & 0.18 & 24.61 \\ 

     CIBP  (0.4) & 21.78 & 44.21 & 25.04 & 6.94 & 1.74 & 0.29 & 26.46  \\ 
  CIBP  (0.5) & 22.46 & 43.10 & 24.44 & 7.82 & 1.87 & 0.31 & 27.91 \\ 

    CRM  & 23.79 & 42.60 & 23.60 & 7.78 & 1.92 & 0.32 & 28.17  \\

     &    \multicolumn{6}{@{}c}{{Scenario 3}}\\
         Toxicity & 10.00 & 15.00 & \textbf{25.00} & 35.00 & 45.00 & 50.00 \\
           CIBP  (0.3) & 3.50 & 22.07 & 47.11 & 23.09 & 3.92 & 0.31 & 22.38 \\ 

           CIBP  (0.4) & 3.37 & 21.87 & 46.68 & 23.73 & 3.97 & 0.38 & 24.31  \\ 
   CIBP (0.5) & 3.54 & 22.51 & 46.14 & 23.47 & 3.93 & 0.41 & 25.67 \\ 

   CRM  & 3.53 & 23.31 & 45.96 & 22.88 & 3.89 & 0.44 & 26.19 \\

     &    \multicolumn{6}{@{}c}{{Scenario 4}}\\
  Toxicity & 5.00 & 10.00 & 15.00 & \textbf{25.00} & 35.00 & 45.00  \\ 
     CIBP  (0.3) & 0.13 & 3.62 & 23.72 & 47.04 & 22.04 & 3.44 & 20.85  \\ 
    CIBP (0.4) & 0.16 & 3.71 & 23.79 & 47.30 & 21.75 & 3.29 & 22.57  \\ 
  CIBP (0.5) & 0.15 & 3.67 & 24.02 & 47.06 & 21.78 & 3.31 & 23.99 \\ 

   CRM  & 0.15 & 3.56 & 24.48 & 47.12 & 21.39 & 3.30 & 24.57 \\ 

     &    \multicolumn{6}{@{}c}{{Scenario 5}}\\
     Toxicity & 2.50 & 5.00 & 10.00 & 15.00 & \textbf{25.00} & 35.00 \\
        CIBP  (0.3) & 0.00 & 0.20 & 4.69 & 25.61 & 45.86 & 23.64 & 19.11 \\ 
 
       CIBP (0.4) & 0.00 & 0.22 & 4.35 & 25.54 & 46.44 & 23.44 & 20.82 \\ 
  CIBP  (0.5) & 0.00 & 0.21 & 4.14 & 25.68 & 46.36 & 23.61 & 22.10  \\ 

   CRM  & 0.00 & 0.23 & 4.09 & 26.50 & 46.24 & 22.94 & 22.45 \\

     &    \multicolumn{6}{@{}c}{{Scenario 6}}\\
 Toxicity & 1.50 & 2.50 & 10.00 & 10.00 & 15.00 & \textbf{25.00}   \\ 
    CIBP (0.3) & 0.00 & 0.03 & 1.74 & 8.63 & 26.63 & 62.97 & 16.33 \\ 

   CIBP (0.4) & 0.00 & 0.03 & 1.33 & 6.53 & 25.96 & 66.16 & 18.01 \\ 
     CIBP (0.5) & 0.00 & 0.02 & 1.24 & 6.20 & 26.59 & 65.94 & 18.88 \\ 

   CRM  & 0.00 & 0.05 & 1.17 & 6.44 & 27.95 & 64.38 & 18.88 \\

\end{tabular}
\end{table}

\end{document}